\documentclass{article}
\usepackage{amsthm,amsmath,amssymb}
\usepackage{graphicx}
\usepackage{fullpage}
\usepackage{color}
\usepackage{hyperref}
\hypersetup{colorlinks=true,linkcolor=blue,filecolor=magenta,urlcolor=cyan,citecolor=blue}
\usepackage{cleveref}
\IfFileExists{dsfont.sty}{\usepackage{dsfont}}{}
\usepackage{blkarray}
\usepackage{booktabs}
\usepackage{caption}
\newtheorem{theorem}{Theorem}
\newtheorem{lemma}[theorem]{Lemma}
\newtheorem{proposition}[theorem]{Proposition}
\newtheorem{observation}[theorem]{Observation}
\newtheorem{defi}[theorem]{Definition}
\newtheorem{corollary}[theorem]{Corollary}

\crefname{observation}{observation}{observations}
\Crefname{observation}{Observation}{Observations}

\allowdisplaybreaks

\newcommand{\Ddc}{D_{\mathrm{dc}}}
\newcommand{\Rdc}{R_{\mathrm{dc}}}
\newcommand{\KING}{\mathrm{KING}}
\newcommand{\MOD}{\mathrm{MOD}}

\title{Directed Cut-Query Complexity of Finding a King}
\author{Ziad Ismaili Alaoui \thanks{University of Liverpool, United Kingdom. \texttt{\{ziad.ismaili-alaoui, nikhil.mande\}@liverpool.ac.uk}}\;\;\thanks{Philipps-Universität Marburg, Germany.}  \and Oded Lachish \thanks{University of London, Birkbeck, United Kingdom. \texttt{o.lachish@bbk.ac.uk} \\ \hspace*{1.8em}Ziad Ismaili Alaoui was funded by EPSRC grant EP/X039447/1.}\and Nikhil S. Mande \footnotemark[1]}

\date{}

\begin{document}

\maketitle
\begin{abstract}
A king in a tournament is a vertex from which every other vertex is reachable by a directed path of length at most two. We study the deterministic query complexity of finding a king when a query specifies a set of vertices and returns the number of edges directed from that set to its complement. We prove that this complexity is $\Theta(n)$ for $n$-vertex tournaments. To show this, we first show that every correct deterministic algorithm in this model must output a maximum-outdegree vertex, even though a king need not have maximum outdegree. We then give an adversary whose answers, after too few queries, remain consistent with a tournament in which the algorithm’s output does not have maximum outdegree.
\end{abstract}

\section{Introduction}
A tournament is an orientation of a complete graph. A vertex is a \emph{king} if it can reach every other vertex by a directed path of length at most two. It is now a well-known fact~\cite{Landau53} that every tournament has a king: any vertex of maximum outdegree is one. Finding such a vertex is a basic search problem when the directions of the edges are initially unknown.

In the standard edge-query model, a query reveals the direction of one edge, and an algorithm's goal is to find a king with as few queries as possible in the worst case. Shen, Sheng, and Wu~\cite{SSW03} proved an $\Omega(n^{4/3})$ lower bound and an $O(n^{3/2})$ upper bound for finding a king. Later work studied finding variants of kings~\cite{BJRS22,LRT22,IAM25}, randomized and quantum algorithms~\cite{MPS23}, and the problem of finding multiple kings~\cite{AGNS24}. Mande, Paraashar, Sanyal, and Saurabh~\cite{MPSS24} studied the communication complexity of finding a king when the edges of the tournament are partitioned between two players.

Cut queries have been extensively studied as a model for solving graph problems without learning the entire graph. In an undirected graph, a cut query specifies a vertex set $S$ and returns the number of edges between $S$ and its complement. The connection to value-oracle submodular function minimization, together with the possibility of solving graph problems without reconstructing the input, has motivated the study of cut-query algorithms for a range of problems~\cite{RSW18, MN20, AEGLMN22, CL23, ASW25, Ken26}.

In the directed cut-query model, a query specifies a vertex set $S$ and returns the number of edges directed from $S$ to its complement. A central question is which problems can be solved using substantially fewer queries than the number of edges. This is particularly relevant for tournaments, which have $\Theta(n^2)$ edges. Ghosh, Kuchlous, Mehra, and Mukhopadhyay~\cite{GKMM26} gave a unified framework for solving tournament problems from the vertex degrees, encompassing earlier results on reachability and strong connectivity and yielding new algorithms for minimum cut. Their framework gives $O(n)$-query algorithms for these problems in the directed cut-query model. We study the directed cut-query complexity of finding a king.

Formally, given a tournament $T$ on the known vertex set $[n]$, a directed cut query specifies a set $S\subseteq[n]$ and returns \[ c_T(S)=|\{(u,v)\in S\times([n]\setminus S):u\to v\text{ in }T\}|. \] Each query has unit cost, and queries may be chosen adaptively. We write $\Ddc(\KING_n)$ for the minimum worst-case number of queries required by a deterministic algorithm that outputs a king of every $n$-vertex tournament.

The following is our main result in this paper.
\begin{theorem}\label{thm:main}
For all sufficiently large $n$,
\[
\frac n3<\Ddc(\KING_n)\le n-1.
\]
\end{theorem}

Our proof in fact gives a lower bound of $(1/2-\epsilon)n$ for every constant $\epsilon>0$ and all sufficiently large $n$.

\subsection{Proof idea}

For the upper bound, we query the outdegrees of $n-1$ vertices, using one singleton cut query for each vertex. Since the sum of all outdegrees is $\binom n2$, we can infer the remaining vertex's outdegree. We then return a maximum-outdegree vertex, which is always a king~\cite{Landau53}.

For the lower bound, we first observe that the answer to any directed cut query is determined by the outdegree of each vertex. Thus an algorithm cannot distinguish two tournaments in which each vertex has the same outdegree in both tournaments. We show that if a vertex does not have maximum outdegree, then the edges can be reoriented, preserving every vertex's outdegree, so that this vertex is no longer a king. Consequently, every correct deterministic cut-query algorithm must output a maximum-outdegree vertex.

It remains to show that finding a maximum-outdegree vertex requires $\Omega(n)$ queries. We show this by an adversarial argument, answering queries as if every vertex had outdegree $(n-1)/2$. Towards a contradiction, suppose a cheap algorithm was correct and outputted vertex $v$ against this adversary. A counting argument shows that, after too few queries, we can perturb the degrees so that they are realized by a tournament consistent with all answers given so far in which another vertex has strictly larger outdegree than that of $v$. The algorithm also outputs $v$ on this tournament, contradicting our earlier conclusion that every correct deterministic cut-query algorithm must output a maximum-outdegree vertex.

\paragraph{Connection with communication complexity.}
Mande, Paraashar, Sanyal, and Saurabh~\cite{MPSS24} proved an $\Omega(n)$ deterministic communication lower bound for finding a king when the edges are partitioned between two players. Since each directed cut query can be simulated using $O(\log n)$ bits of communication, this implies an $\Omega(n/\log n)$ lower bound on the directed cut-query complexity of finding a king. They also proved an $\Omega(n\log n)$ deterministic communication lower bound for finding a maximum-outdegree vertex. Our result that every correct deterministic cut-query algorithm must output a maximum-outdegree vertex therefore gives an alternative proof of the $\Omega(n)$ lower bound. We give the details in \Cref{sec:communication}.

\section{Equivalence of king-finding and maximum-outdegree search}\label{sec:structure}

We first show that directed cut queries reveal no information beyond the outdegree of each vertex. For a tournament $T$ and a vertex $v$, let $d_T^+(v)$ denote the outdegree of $v$ in $T$.

\begin{lemma}\label{lem:indistinguishable}
Let $T_1$ and $T_2$ be tournaments on the same labelled vertex set $V$, with $d_{T_1}^+(v)=d_{T_2}^+(v)$ for every $v\in V$. Then every deterministic directed cut-query algorithm has the same execution and output on $T_1$ and $T_2$.
\end{lemma}

\begin{proof}
The sum $\sum_{v\in S}d_T^+(v)$ counts every edge whose origin lies in $S$. These are precisely the edges directed from $S$ to its complement and the edges within $S$, each counted once. Since $T$ is a tournament, there are exactly $\binom{|S|}{2}$ edges within $S$. Therefore,
\begin{equation}\label{eq:cut}
c_T(S)=\sum_{v\in S}d_T^+(v)-\binom{|S|}{2}.
\end{equation}
Since $d_{T_1}^+(v)=d_{T_2}^+(v)$ for every $v\in V$, the right-hand side is the same for $T_1$ and $T_2$ for every choice of $S$. Hence $c_{T_1}(S)=c_{T_2}(S)$ for every $S\subseteq V$. Since the algorithm is deterministic, it asks the same queries and produces the same output on both tournaments.
\end{proof}

For a tournament $T$ on vertex set $[n]$, we write
\[
 {\bf d}(T)=(d_T^+(1),\ldots,d_T^+(n)).
\]
This is a \emph{labelled} vector: its $i$th coordinate belongs to vertex $i$. Equality of vectors always means coordinatewise equality, not equality up to permutation. A vector is \emph{realizable} if it equals ${\bf d}(T)$ for some tournament $T$ on $[n]$. We write $N_T^+(v)$ for the out-neighbours of $v$, and $N_T^+[v]=N_T^+(v)\cup\{v\}$. Subscripts are omitted when the tournament is clear. 

\begin{defi}
We say that a vertex $w$ \emph{covers} $v$ if $w\to v$ and $w\to x$ for every $x\in N^+(v)$.
\end{defi}

We make the simple observation that if a vertex is covered, then it cannot be a king. This is because if $w$ covers $v$, then there is no path of length two from $v$ to $w$.

\begin{observation}\label{obs:cover-no-king}
    If a vertex $w$ covers another vertex $v$ in a tournament $T$, then $v$ is not a king in $T$.
\end{observation}

Conversely, if $v$ is not a king, choose a vertex $w$ that it cannot reach in at most two steps. Then $w$ covers $v$. In particular, if $w$ covers $v$, then $d^+(w)\ge d^+(v)+1$. Consequently,
\begin{equation}\label{eq:cover-degree}
v \textnormal{ is not a king } \implies \exists w \textnormal{ such that } d^+(w)\ge d^+(v)+1.
\end{equation}
In particular, the contrapositive of the above recovers the fact that every maximum-outdegree vertex is a king~\cite{Landau53}.

We next show that any vertex whose outdegree is not maximum can be made a non-king by reorienting edges while preserving every vertex's outdegree. We will use the following observation.
\begin{observation}\label{obs:triangle}
For any directed triangle in a tournament, reversing all of its three edges preserves the outdegree of every vertex.
\end{observation}
\begin{proof}
Each vertex of the triangle has exactly one outgoing edge within the triangle both before and after the reversal. All other edges are unchanged.
\end{proof}

\begin{lemma}\label{lem:cover}
Let $\mathbf d=(d_1,\ldots,d_n)$ be a realizable degree vector, and let $v,w\in[n]$ satisfy $d_w>d_v$. There is a tournament $T$ with $\mathbf d(T)=\mathbf d$ in which $w$ covers $v$.
\end{lemma}

\begin{proof}
Start with any realization $T$ of $\mathbf d$. We first arrange that $w\to v$. If instead $v\to w$, there is a vertex $y$ with $w\to y\to v$: otherwise $N^+(w)\subseteq N^+(v)$, contradicting $d_w>d_v$. Reverse the directed triangle $v\to w\to y\to v$. By \Cref{obs:triangle}, the degrees are preserved and we now have the edge $w\to v$. With $w\to v$ fixed, let
\[
A=N^+(v)\setminus N^+(w),\qquad B=N^+(w)\setminus\bigl(N^+(v)\cup\{v\}\bigr).
\]
Writing $C=N^+(v)\cap N^+(w)$, we have $d_v=|C|+|A|$ and $d_w=|C|+|B|+1$. Hence 
\begin{equation}\label{eq:ba-geq-0}
|B|-|A|=d_w-d_v-1 \ge 0.
\end{equation}
If $A$ is nonempty, then $B$ is also nonempty, so we may choose $x\in A$ and $y\in B$. The definitions give $v\to x\to w\to y\to v$. There are two cases to consider:

\begin{itemize}
    \item If $x\to y$, reverse the triangle $v\to x\to y\to v$. Vertex $v$ loses $x$ and gains $y$ as an out-neighbour, while $N^+(w)$ is unchanged. Since $y\in N^+(w)$, the new set $A$ is exactly $A\setminus\{x\}$.
    \item If $y\to x$, reverse the triangle $w\to y\to x\to w$. Vertex $w$ loses $y$ and gains $x$ as an out-neighbour, while $N^+(v)$ is unchanged. Since $y\notin N^+(v)$, again the new set $A$ is exactly $A\setminus\{x\}$.
\end{itemize}

In both cases every degree is preserved, the edge $w\to v$ is unchanged, and $|A|$ decreases by one. Redefining $A,B,C$ using the updated tournament, we again have $d_v=|C|+|A|$ and $d_w=|C|+|B|+1$. Thus $|B|-|A|=d_w-d_v-1\ge0$ by \Cref{eq:ba-geq-0}, so whenever $A$ is nonempty, $B$ is also nonempty and another step is possible. This process therefore terminates with $A=\emptyset$, giving $N^+(v)\subseteq N^+(w)$. Together with $w\to v$, this means that $w$ covers $v$, and the resulting tournament still satisfies $\mathbf d(T)=\mathbf d$.
\end{proof}

\begin{corollary}\label{cor:universal}
For every realizable degree vector $\bf d$, the vertices that are kings in every realization of $\bf d$ are exactly the vertices of maximum outdegree in $\bf d$.
\end{corollary}
\begin{proof}
A maximum-outdegree vertex is a king by \Cref{eq:cover-degree}. Every other vertex is covered in some realization by \Cref{lem:cover}, and hence not a king by \Cref{obs:cover-no-king}.
\end{proof}

Let $\MOD_n$ denote the task of finding a maximum-outdegree vertex.

\begin{corollary}\label{cor:equivalence}
Every correct deterministic directed cut-query algorithm for $\KING_n$ outputs a maximum-outdegree vertex on every input. Consequently,
\[
 \Ddc(\KING_n)=\Ddc(\MOD_n).
\]
\end{corollary}
\begin{proof}
Let $A$ be a correct deterministic directed cut-query algorithm for $\KING_n$. Suppose that on some input tournament $T$, it outputs a vertex $v$ that does not have maximum outdegree. By \Cref{cor:universal}, there exists a tournament $T'$ with $\mathbf d(T')=\mathbf d(T)$ such that $v$ is not a king in $T'$. By \Cref{lem:indistinguishable}, the output of $A$ on $T'$ is the same as its output on $T$, namely $v$. This contradicts the correctness of $A$.

Thus every correct deterministic algorithm for $\KING_n$ also solves $\MOD_n$, which implies $\Ddc(\KING_n)\ge\Ddc(\MOD_n)$. The reverse inequality follows because every maximum-outdegree vertex is a king by \Cref{eq:cover-degree}.
\end{proof}

\begin{proposition}\label{prop:upper}
For every $n\ge1$, $\Ddc(\KING_n)\le n-1$.
\end{proposition}
\begin{proof}
Query $c_T(\{v\})=d_T^+(v)$ for $n-1$ vertices. Infer the remaining degree from $\sum_v d_T^+(v)=\binom n2$, and output a vertex of maximum outdegree.
\end{proof}

\section{A direct lower bound}\label{sec:lower}

In this section we first note sufficient conditions for a degree vector to be realizable, and then give the counting argument used to preserve the query answers.

\subsection{Realizing bounded degree perturbations}
\begin{theorem}[Landau~\cite{Landau53}, {\cite[Theorem~29]{Moon68}}]\label{thm:landau}
Let $d_1\le\cdots\le d_n$ be nonnegative integers. The vector ${\bf d} = (d_1, \dots, d_n)$ is realizable if and only if
\[
 \sum_{i=1}^k d_i\ge\binom k2\quad(1\le k\le n),
 \qquad
 \sum_{i=1}^n d_i=\binom n2.
\]
\end{theorem}

\begin{lemma}\label{lem:middle}
Let $n\ge4$ and ${\bf d}\in\mathbb Z^n$ satisfy
\[
 \frac n4\le d_i\le\frac{3n}{4}\quad(i\in[n]),
 \qquad \sum_{i=1}^n d_i=\binom n2.
\]
Then $\bf d$ is realizable.
\end{lemma}
\begin{proof}
Sort the coordinates into nondecreasing order. For $k\le n/2+1$,
\[
 \sum_{i=1}^k d_i\ge\frac{nk}{4}\ge\frac{k(k-1)}2 = \binom{k}{2}.
\]
For $k\ge n/2+1$, the bound on the remaining coordinates gives
\begin{align*}
 \sum_{i=1}^k d_i-\binom k2 & = \sum_{i = 1}^n d_i - \sum_{i = k+1}^nd_i - \binom{k}{2} \ge \binom n2-\frac{3n}{4}(n-k)-\binom k2 \\
 & = \frac{n(n-1) - k(k-1)}{2} - \frac{3n}{4}(n-k)\\
 & = \frac{(n-k)(n+k-1)}{2} - \frac{3n}{4}(n-k) = \frac{(n-k)(2k-n-2)}4\ge0,
\end{align*}
where the last inequality holds since we are in the case where $k \geq n/2 + 1$.
Thus the conditions in \Cref{thm:landau} are satisfied, implying a realization of the sorted vector, and hence the original vector by relabeling vertices.
\end{proof}

\subsection{An integer vector preserving all answers}

\begin{lemma}\label{lem:pigeonhole}
Let $M\in\{0,1\}^{k\times n}$ and let $R\ge1$ be an integer. If
\[
 (R+1)^n>(nR+1)^k,
\]
then there is a nonzero $z\in\mathbb Z^n$ such that $Mz=0$ and $|z_i|\le R$ for every $i\in[n]$.
\end{lemma}
\begin{proof}
There are $(R+1)^n$ vectors $x\in\{0,1,\ldots,R\}^n$. Each coordinate of $Mx$ is a sum of at most $n$ coordinates of $x$, since every entry of $M$ is either $0$ or $1$. Each summand is an integer between $0$ and $R$, so each coordinate of $Mx$ is an integer between $0$ and $nR$. Each of the $k$ coordinates of $Mx$ has at most $nR+1$ possible values, namely $0,1,\dots,nR$. Hence, as $x$ ranges over $\{0,1,\dots,R\}^n$, the vector $Mx$ takes at most $(nR+1)^k$ distinct values. By our assumption and the pigeonhole principle, there exist two distinct vectors $x,y \in \{0, 1, \dots, R\}^n$ with the same image. Set $z=x-y$. Since $x\ne y$, we have $z\ne0$. Moreover, $Mz=Mx-My=0$, and $|z_i|\le R$ for every $i\in[n]$.
\end{proof}

We now combine the preceding two lemmas to prove the lower bound. We answer queries according to a regular tournament and show that, after too few queries, there is a realizable degree vector consistent with all answers in which the algorithm's output does not have maximum outdegree.
\begin{theorem}\label{thm:odd}
Let $n\ge17$ be odd. Then
\[
\Ddc(\KING_n)\ge \frac n2-\frac{3n}{2\log n}-1.
\]
\end{theorem}
\begin{proof}
For this proof, fix $m = (n-1)/2$ and $R = \lfloor n/8 \rfloor$.
Fix a correct deterministic algorithm and run it on an $n$-vertex regular tournament, that is, one in which every vertex has outdegree $m$. It is well known that such a tournament exists for every odd $n$~\cite{Moon68}.

Suppose this execution makes $q$ queries $S_1,\ldots,S_q$ and outputs vertex $v$. Define $M\in\{0,1\}^{(q+1)\times n}$ by
\[
 M_{1,j}=1,\qquad M_{i+1,j}=\begin{cases}1&j\in S_i,\\0&j\notin S_i,\end{cases}
 \quad i\in[q],\ j\in[n].
\]
Assume, towards a contradiction, that
\begin{equation}\label{eq:count-condition}
 (R+1)^n>(nR+1)^{q+1}.
\end{equation}
By \Cref{lem:pigeonhole}, there is a nonzero integer vector $z$ with $Mz=0$ and $|z_i|\le R$. The first row gives $\sum_i z_i=0$, and the remaining rows give
\begin{equation}\label{eq:query-zero}
 \sum_{j\in S_i}z_j=0\qquad(i\in[q]).
\end{equation}
In particular, $z$ is not constant. We may replace $z$ by $-z$ so that $z_v$ is not a maximum coordinate of $z$: if $z_v$ is already nonmaximum, keep $z$; otherwise choose $w$ with $z_w<z_v$ and negate $z$.

Set $d_i=m+z_i = (n-1)/2 + z_i$ for every $i \in [n]$. Then ${\bf d} = (d_1, \dots, d_n)$ is an integer vector with total sum $n(n-1)/2 = \binom n2$. Moreover, \[
 m-R\ge\frac{3n}{8}-\frac12\ge\frac n4,
 \qquad
 m+R\le\frac{5n}{8}-\frac12\le\frac{3n}{4}.
\]
Thus, each $d_i \in [n/4, 3n/4]$.
By \Cref{lem:middle}, $\bf d$ is realizable. Choose $w$ with $z_w>z_v$. By \Cref{lem:cover}, there is a realization $T'$ of $\bf d$ in which $w$ covers $v$.

For each queried set $S_i$, the sum of the outdegrees in $T'$ is
\[
\sum_{j\in S_i}d_j=\sum_{j\in S_i}(m+z_j)=m|S_i|+\sum_{j\in S_i}z_j=m|S_i|,
\]
where the last equality follows from \Cref{eq:query-zero}. This is the same degree sum as in the regular tournament. By \Cref{eq:cut}, the answer to the query $S_i$ is therefore the same on both tournaments. This holds true for all $i \in [q]$. The algorithm therefore follows the same execution on $T'$ and outputs $v$. This is incorrect because $w$ covers $v$ in $T'$ and hence $v$ cannot be a maximum-outdegree vertex or king in $T'$ by \Cref{obs:cover-no-king}.

Thus \Cref{eq:count-condition} is impossible. Taking logs and cross-multiplying gives
\[
 q+1\ge\frac{n\log(R+1)}{\log(nR+1)}.
\]
Since $R=\lfloor n/8\rfloor$, we have $R+1\ge n/8$ and $nR+1\le n^2$. Therefore,
\[
q+1\ge\frac{n\log(R+1)}{\log(nR+1)}\ge\frac{n(\log n-3)}{2\log n}=\frac n2-\frac{3n}{2\log n}.
\]
Thus $q\ge n/2-3n/(2\log n)-1$, proving the claimed lower bound.
\end{proof}

\subsection{Proof of main result}

\begin{lemma}\label{lem:padding}
For every $n\ge2$, $\Ddc(\KING_n)\ge\Ddc(\KING_{n-1})$.
\end{lemma}
\begin{proof}
Given a tournament $T$ on $[n-1]$, add a new vertex $s$ and direct every edge incident to $s$ towards $s$, obtaining $T'$. For a query $S\subseteq[n-1]\cup\{s\}$, put $A=S\cap[n-1]$. Then
\[
 c_{T'}(S)=c_T(A)+\begin{cases}|A|&s\notin S,\\0&s\in S.\end{cases}
\]
Thus one query to $T$ suffices to simulate a query to $T'$. The vertex $s$ cannot be a king of $T'$, and it cannot be an internal vertex of a directed path. Consequently, every king of $T'$ is a king of $T$. An algorithm for $\KING_n$ therefore yields one for $\KING_{n-1}$ with no more queries.
\end{proof}

\begin{proof}[Proof of \Cref{thm:main}]
The upper bound is \Cref{prop:upper}. Let $m$ be the largest odd integer at most $n$. By \Cref{lem:padding} and \Cref{thm:odd}, for all sufficiently large $n$,
\[
\Ddc(\KING_n)\ge\Ddc(\KING_m)\ge\frac m2-\frac{3m}{2\log m}-1.
\]
Since $m\ge n-1$, the right-hand side is at least $(1/2-\epsilon)n$ for every fixed $\epsilon>0$ and all sufficiently large $n$. In particular, it exceeds $n/3$ for all sufficiently large $n$.
\end{proof}

\section{An alternative lower bound via communication}\label{sec:communication}
The structural equivalence in \Cref{cor:equivalence} also allows us to deduce an $\Omega(n)$ lower bound from the deterministic maximum-outdegree communication lower bound of $\Omega(n \log n)$ of Mande et al.~\cite[Theorem~5]{MPSS24}.

Fix a publicly known partition of the unordered vertex pairs between Alice and Bob. Each player knows the directions of the edges assigned to them. We now show how Alice and Bob can simulate a $q$-cost cut-query algorithm by an $O(q \log n)$-cost communication protocol. To simulate a directed cut query $S$, each player counts the edges they own directed from $S$ to its complement and sends this count to the other player. Each count is at most $\binom n2$, so this can be sent to each other using $O(\log n)$ bits of communication. Both players then know the answer to the cut-query $S$ and can determine the next query of the deterministic cut-query algorithm without further communication. Since the cut-query algorithm makes at most $q$ queries, the cost of the communication protocol is $O(q \log n)$.

By \Cref{cor:equivalence}, a correct cut-query algorithm for finding a king must output a maximum-outdegree vertex on every tournament. The $\Omega(n\log n)$ communication lower bound for a suitable partition therefore implies $q=\Omega(n)$.

\section{Further directions}
An interesting special case is finding a king when the input is promised to be a transitive tournament. Since its outdegrees form a permutation of $\{0,\ldots,n-1\}$ and its unique king is its source, \Cref{eq:cut} makes this equivalent to finding the location of $n$ in an unknown permutation of $[n]$, where each query returns the sum of an arbitrary subset of entries. Determining the query complexity of this problem seems to be an independently interesting puzzle.

Another natural next question is to determine the \emph{randomized} directed cut-query complexity of finding a king. Let $\Rdc(\KING_n)$ denote the minimum worst-case query bound of an algorithm that outputs a king with probability at least $2/3$ on every input. The deterministic algorithm from \Cref{prop:upper} gives $\Rdc(\KING_n)\le n-1$. The randomized $\Omega(n)$ communication lower bound for finding a king~\cite[Theorem~4]{MPSS24}, together with the simulation in \Cref{sec:communication} (which can also easily be seen to work in the randomized setting), gives
\[
 \Omega\!\left(\frac n{\log n}\right)\le\Rdc(\KING_n)\le n-1.
\]
Does randomization permit an $o(n)$-cost cut-query algorithm to find a king, or does the linear lower bound persist?

Our structural equivalence between $\KING_n$ and $\MOD_n$ in \Cref{cor:equivalence} relies on correctness on every input. For each nonmaximum vertex, \Cref{lem:cover} gives a realization in which that vertex is not a king, but this realization may depend on the vertex. Moreover, the tournament constructed in the proof of \Cref{thm:odd} depends on the queries made by the algorithm. These arguments therefore do not directly yield a matching randomized lower bound via Yao's principle, which requires a single hard input distribution against all deterministic algorithms with a given query budget.
\bibliography{bib}

\end{document}